\documentclass[prd,superscriptaddress,unsortedaddress,twocolumn,showpacs,preprintnumbers,amsmath,amssymb,floatfix]{revtex4-1}
\usepackage[colorlinks=true, pdfstartview=FitV, linkcolor=red, citecolor=blue, urlcolor=blue]{hyperref}

\usepackage{graphicx}
\usepackage{bm}

\newcommand\beq{\begin{equation}}
\newcommand\eeq{\end{equation}}

\usepackage{amsmath} 

\begin{document}

%\preprint{RBRC-TBA}
\title{Quark contribution for center domain in heavy ion collisions}

\author{Kouji Kashiwa}
\email[]{kashiwa@ribf.riken.jp}
\affiliation{RIKEN BNL Research Center, Brookhaven National Laboratory, Upton, NY 11973}

\author{Akihiko Monnai}
\email[]{amonnai@riken.jp}
\affiliation{RIKEN BNL Research Center, Brookhaven National Laboratory, Upton, NY 11973}

%\date{\today}

\begin{abstract}
The center domain structure is revisited with introduction of quark
contribution to understand phenomenology in high-energy heavy ion collisions.
We show the quark contribution may allow metastable states that would lead to a natural and consistent
explanation for the temperature dependences of color opacity and viscosity.
We also argue the possibility of indirect observations of the center
domain structure in experiments due to a topological critical temperature.
\end{abstract}

\maketitle

%%%%%%%%%%%%%%%%%%%%%%%%%%%%%%%%%%%%%%%%%%%%%%%%%%%%%%%%%%%%%%%%%%%%%%%%
% Paper
%%%%%%%%%%%%%%%%%%%%%%%%%%%%%%%%%%%%%%%%%%%%%%%%%%%%%%%%%%%%%%%%%%%%%%%%

It is almost infallibly considered that recent experimental programs at
the Relativistic Heavy Ion Collider (RHIC) and the Large Hadron Collider
(LHC) achieve the materialization of the quark-gluon plasma (QGP) \cite{Yagi:2005yb}.
Experimental data suggest the hot medium created in those nucleus-necleus collisions is characterized by large color
opacity and near perfect fluidity.
The former is quantified by suppression and energy redistribution of
jets traveling through the medium called jet quenching
\cite{Adler:2002tq}. The phenomenon is considered to be one of the
strong evidences for the existence of a hot medium in heavy ion
collisions.
The latter, on the other hand, is supported by the existence large
azimuthal momentum anisotropy in hadronic particle spectra, which is
well-described in relativistic hydrodynamic pictures
\cite{Schenke:2010rr}. This indicates the QGP is strongly coupled in the
vicinity of the quark-hadron crossover temperature, as opposed to the
weakly-coupled picture one would na\"{i}vely expect from perturbative quantum
chromodynamics (pQCD). It is now widely-accepted that its viscosity is smaller than that of any matter we have found so far -- it is close to the lower bound conjectured in kinetic theory \cite{Danielewicz:1984ww} and Anti-de Sitter/conformal field theory
correspondence \cite{Kovtun:2004de}. Therefore it would be of great
importance to establish microscopic explanations for the mechanisms
behind the parton energy loss and the small viscosity of the hot medium.

Recently, those properties of the QGP are shown to be qualitatively
understood from the center domain structure of the deconfined matter
\cite{Asakawa:2012yv}. The jet quenching is interpreted as reflecting
and scattering of partons by center domain walls. The short mean free
path is characterized by the typical size of domains.
In this letter, we consider the quark contribution to the center domain
structure and explore its possible consequences in the phenomenology of the QGP in experiments. We find there may exist a critical temperature
that separates the strongly-coupled and the perturbative pictures.

The reached initial temperature at RHIC is roughly estimated as 
$300$-$600$ MeV \cite{Adare:2008ab} and that at LHC is indicated to be even higher \cite{Wilde:2012wc}. A strongly-coupled QGP (sQGP) is expected to be realized in those heavy ion collisions.
In such temperature regime, the center domain structure may appear
\cite{Asakawa:2012yv}. The color glass condensate \cite{McLerran:1993ni} and subsequent glasma \cite{Lappi:2006fp} pictures imply the
correlation length in the transverse plane is around $\sim
1/Q_s$. Thus the quark matter can develop ``domains" with different
phases in the early stage of the collision dynamics, which should
survive to some extent in later stages after time-evolution as well \cite{Mohapatra:2012ck} and characterize the
sQGP.
When there are no flavors, ${\mathbb Z}_3$ (center) symmetry can be realized in the
system. On the other hand, the existence of quarks can explicitly break
the symmetry and might allow \textit{metastable states}. In classical
non-equilibrium thermodynamics, metastable states are keys in explaining
phenomena such as supercooling and supersaturation. Thus in analogy one
might be able to expect a new heavy ion phenomenology to arise from this
topological supercooling of the QGP.

To understand the center domain structure, we consider the perturbative
one-loop effective potential of the gluon and the massless quark here.
Our interest is sQGP realized in RHIC and LHC and thus we may neglect
the chiral symmetry breaking.
By considering the expansion about a background field for the time-like
component of the vector potential
\begin{align}
(A_4^{cl})^{ab} &= \frac{2 \pi T}{g} q_a \delta^{ab};
\end{align}
$a$ and $b$ are color indices where $T$ is temperature and $g$ means the
gauge coupling,
the gluon and quark perturbative one-loop effective potentials
\cite{Gross:1981br,Weiss:1980rj}
are expressed as
\begin{align}
F_g
&= \frac{2 \pi^2 T^4}{3}
   \sum_{a,b} \Bigl( 1 - \frac{\delta_{ab}}{3} \Bigr) B_4 (|q_a-q_b|_\mathrm{mod\ 1}) ,
\\
F_f
&= - \frac{4 \pi^2 N_f T^4}{3}
   \sum_{a} B_4 \Bigl( \Bigl| q_a+\frac{1}{2} \Bigr|_\mathrm{mod\ 1} \Bigr) ,
\end{align}
with the forth Bernoulli polynomial
\begin{align}
B_4(x) &= x^4 -2x^3 + x^2 - \frac{1}{30},
\end{align}
where $N_f$ is the flavor number.
In this background field, the Polyakov-loop in the fundamental
representation can be expressed as
\begin{align}
\Phi
&= \frac{1}{3} \mathrm{tr} {\cal P}
   ~e^{ig \int_0^{1/T}A_4 (\tau,{\vec x})~d\tau }
 = \frac{1}{3}
   \sum_{a=1}^3 e^{2\pi i q_a},
\end{align}
where $\sum_a q_a=0$ because $A_4$ is an element of the $SU(3)$ Lie
algebra.
The Polyakov-loop at minima of the effective potential
$F = F_g + F_f$
in the high $T$ limit can
be expressed as $\Phi = \exp ( 2 \pi i \nu / 3 )$
where $\nu = 0,1$ and $2$.
If we discuss the non-perturbative effects, we need a model of QCD.

\begin{figure}[tbp]%[H]
\begin{center}
 \includegraphics[width=0.35\textwidth]{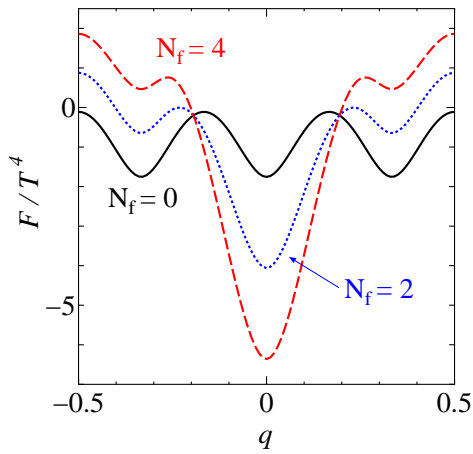}
\end{center}
\caption{ The flavor-dependence of the effective potential normalized by
 $T^4$.
}
\label{Fig:EP-Nf}
\end{figure}

The flavor-dependence of the effective potential normalized by $T^4$ is
shown in Fig.~\ref{Fig:EP-Nf} where $q_i$ are set as $q_1 = q_2 = q$.
%%%%%%%%%%%%%%%% Fig %%%%%%%%%%%%%%%%%%%%%
%%%%%%%%%%%%%%%%%%%%%%%%%%%%%%%%%%%%%%%%%%
This figure was already shown in Ref.~\cite{Belyaev:1991np}.
The existence and its physical meaning of metastable minima are still
controversial~\cite{Belyaev:1991np,Smilga:1993vb}.
In recent lattice QCD
simulation~\cite{Borsanyi:2010cw,Deka:2010bc,Danzer:2010ge},
the center domain with fundamental quark contribution were
investigated and then some evidence were obtained.
It should be noted that description of such metastable mimima
is difficult from the viewpoint of the standard (equilibrium)
thermodynamics because the free energy should be a convex
function \cite{Lieb:1997mi}.
Also, nonzero $q_i$ acts as the imaginary chemical potential and thus it
may leads the nonzero pure imaginary quark number density at zero
chemical potential at metastable minima and then
the interpretation of the pure imaginary quark number
is not obvious.
These issues may be related with the incompleteness of theory which can
describe thermodynamics of the non-equilibrium system.
In this letter,
we treat the metastable minima of the effective potential are
actual metastable states in non-equilibrium system and then
the center domain is a physical object.
In the end of this letter, we show the qualitative difference of
experimental data with and without the center domain structure.
If it will be observed, the center domain can be considered as the physical
object and then it may provide useful information to construct the theory
of the non-equilibrium system.

From Fig.~\ref{Fig:EP-Nf},
we can see that the energy barrier from $\nu=1, 2$ to $0$ becomes lower
when the flavor number increases. On the other hand, the energy barrier
from $\nu=0$ to $1,2$ becomes higher. It should be noted that metastable
minima would become unphysical above $N_f \sim 3$ when they reach the
$F>0$ region because their pressures become negative and thermodynamics
is ill-defined. It is also note-worthy that non-colored particles, such as
leptons, can have positive pressure even in such systems so that the
``overall pressure" is positive. However, the metastable states are
still forbidden because their interactions with quarks and gluons are
weak and the combined system should not be considered as a thermalized
medium. This implies one can define a critical temperature
$T^\mathrm{cri}$ from the thermodynamic instability of the metastable minima that
characterize a major change in the center domain structure.

The quark contribution would be essential in providing a bridge from the hydrodynamic picture, where the mean free path $\lambda$ is short, to the pQCD picture where it is very long. Below $T^\mathrm{cri}$, the mean free path in the quark matter is characterized by the effective size of the domain. One can expect that $\lambda$ would gradually increase with the number of flavors, because if the pressures of $\nu = 1,2$ domains decrease as shown in Fig.~\ref{Fig:EP-Nf}, the pressure balance requires the expansion of $\nu = 0$ domain. This indicates a longer mean free path since when the quark contribution is not very large, it is given as
\begin{align}
\lambda \simeq 3^{1/2} R_d \sum_j \bigg( \frac{V_j}{\sum_i V_i} \bigg)^{3/2}, 
\label{eq:lambda}
\end{align}
where $R_d$ is the domain size in the ${\mathbb Z}_3$ symmetric case and $V_i$ is the volume portion of a domain $\nu = i$ ($i,j=0,1,2$) when the pressures are equal. The minimum is given when $V_0 = V_1 = V_2$, which can be realized when all the domains have the same pressure, \textit{i.e.}, in the pure gauge case. It is note-worthy that the energy barrier towards the $\nu = 0$ domain could be decreased as well. Considering the typical initial size of individual domains $\sim Q_s^{-1}$ tends to shrink for larger energies, the effect would be important in understanding the latest experimental implications for the RHIC and LHC collisions in hydrodynamic analyses that shear viscosity would increase with the temperature from $(\eta/s)_\textrm{RHIC} \simeq 0.12$ to $(\eta/s)_\textrm{LHC} \simeq 0.2$ \cite{Gale:2012rq}. The probability for realization of $\nu = 0$ state itself becomes larger, inducing the merging of neighboring domains. This would lead to rapid increase in the mean free path near $T^\mathrm{cri}$. At $T^\mathrm{cri}$, one might observe a weakly-coupled QGP (wQGP) because there could still be the energy barrier between the domains just below the temperature, but such barrier no longer exists slightly above $T^\mathrm{cri}$. Since the effective number of flavors is already $N_f \sim$ 2-3 for the energy scales of the recent heavy ion experiments, depending on the precise form of the effective potential a very beginning of such behavior might well be observed in future heavy ion experiments if the center domain structure plays an essential role in the QGP physics. 

\begin{figure}[tbp]%[H]
\begin{center}
 \includegraphics[width=0.44\textwidth]{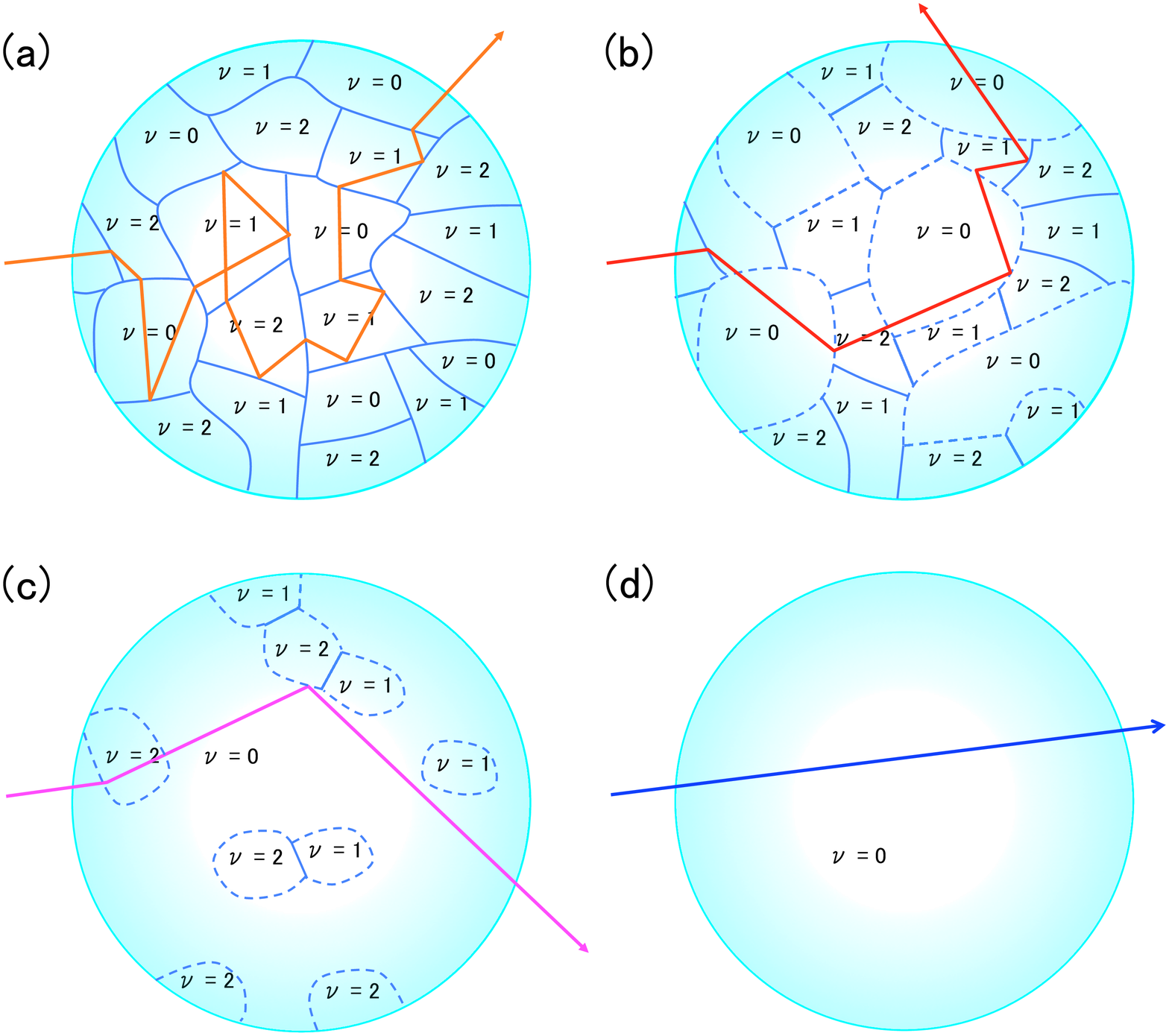}
\end{center}
\caption{ Possible schematic pictures of the temperature-dependence of parton scattering by broken center domains. Dashed line denotes $\nu = 0$ to/form $1,2$ boundary with quark contribution. (a) Mean free path is typically of the domain size in the pure gauge scenario \cite{Asakawa:2012yv}. (b) Domains are partially broken due to metastable minima induced by quarks. (c) $\nu = 0$ is significantly favored at higher energies and those domains connect to form large clusters. $\nu = 1,2$ shrinks as their pressures decrease. (d) The metastable states vanish above $T^\mathrm{cri}$ before their pressures become negative.
}
\label{Fig:domain}
\end{figure}

The schematic figures of the flavor/temperature dependence of the topological domain structure are shown in Fig.~\ref{Fig:domain} (a)-(d). When 
there is no quark contribution, the center domains are stable and the
mean free path of a parton is characterized by their size. When there is
a moderate number of flavors, on the other hand, the stable domains
($\nu = 0$) expands while the metastable ones ($\nu = 1,2$) shrink. The energy barrier of domain walls are also reduced for the patrons traveling from the metastable
to the stable states, allowing longer effective mean free path. As the
number of flavors increase with the temperature, $\nu = 0$ domain is
more significantly favored and the domains starts to form large clusters as
topological percolation occurs. Note this starts to happen even when the system does not
reach $T^\mathrm{cri}$. Finally the metastable states vanish above the
critical temperature, allowing very long mean free path. The rapid
change in the center domain structure could be observed as a sudden
reduction in jet quenching and increase in viscosity at sufficiently high temperatures.

%%%%%%%%%%%%%%%% Fig %%%%%%%%%%%%%%%%%%%%%
%%%%%%%%%%%%%%%%%%%%%%%%%%%%%%%%%%%%%%%%%%

Jet quenching is characterized with the parton energy loss par unit length $dE/dx$. 
The energy loss by the center domains is obviously reduced for a smaller number of walls. On the other hand, the energy barrier from stable to metastable states increases considerably, leading to enhancement of the energy loss. The competition can lead to non-trivial energy dependence, which requires further investigation with more quantitative analyses. The fact increased suppression is observed for mid-low momentum region in higher energy collisions \cite{CMS:2012aa} implies the energy barrier effect is stronger at RHIC and LHC energies. At a sufficiently high temperature, those structure disappears and one should observe much more color-transparent plasma in the low-momentum region. 
High momentum behavior is considered to be described with pQCD as encouraged by experimental data at RHIC and LHC, possibly with running coupling corrections.

When the initial temperature is above $T^\mathrm{cri}$, there would be no domain wall structure. Therefore, if the center domains are the dominant reason behind the collective properties of the QGP, heavy ion collisions of higher energies could be very unique. At the initial stage, a wQGP would be produced with only $\nu = 0$ domain. As the temperature decreases with time-evolution of the expanding system, metastable states are thermodynamically allowed but not chosen since all the domains are already in the true minimum state. The QGP would be gas-like and develop very small elliptic flow at most in the magnitude expected from pQCD during the evolution until the hadronic freeze-out. On the other hand, if the center domains are not the main reason behind the fluidity, wQGP could smoothly transit into sQGP as the system cools down and develop a sizable elliptic flow. This implies that if a sudden change in elliptic flow coefficient is observed, that could be interpreted as a sign of breaking of topological objects in the color plasma.

The typical size of a domain would be $R_d \sim 0.5$ fm
\cite{Asakawa:2012yv} while the radius of a proton is $\sim 0.8$
fm. This implies that a very few number of center domain structures could be
formed even in $p$-$p$ collisions. Most of them, however, could merge
into one domain as the quark contribution encourages the $\nu = 0$
state, losing fluidity and color opacity of the system. One may observe an onset of collective behavior when the transverse volume
size of a hot medium becomes larger in $p$-$p$ and $p$-$A$ collisions at  much higher energies. It should be noted that one also has to be careful since low-temperature boundary effects would be non-negligible and can break the domain wall structures at initial stages in such collisions.

It is important to keep note that the center domain structure could naturally arise by assuming the existence of color flux tube spots on the transverse plane, which is an implication of color glass theory. The domain structure in the longitudinal direction, on the other hand, could be more non-trivial. If the flux tube survives throughout the collision processes, it could support highly-anisotropic hydrodynamic pictures \cite{Martinez:2010sc}. If fragmentation of the flux tubes takes place during the expansion of the heavy ion system, the longitudinal domain size would become finite and three-dimensional fluidity and opacity may be achieved. The fact that no directional preferences are observed in the data of jet quenching might favor the latter scenario.

In summary, we have explored the center domain structure in the QGP with the quark contribution and its implications in heavy ion physics. We show the flavor effects can accommodate correct behavior of viscosity and jet quenching from RHIC to LHC. The picture may pose a new critical temperature that can separate strongly-coupled and weakly-coupled QGPs by the break-down of metastable center domains. If sudden changes in the observables are detected, that could be the first time a topological object is observed in experiments. Since the number of effective flavors is already close to the critical value, a hint of such phenomena could be observed in future heavy ion experiments with higher energies, opening a brand new physics of LHC just as the quark-hadron transition at RHIC did a decade ago if center domains with broken ${\mathbb Z}_3$ symmetry are present in the QGP.

\noindent
\begin{acknowledgments}
The authors are grateful for encouragements and comments by R. Pisarski and R. Venugopalan. 
This work is supported by RIKEN Special Postdoctoral Researchers Program.
\end{acknowledgments}

\bibliography{ref.bib}

\begin{thebibliography}{99}

%\cite{Yagi:2005yb}
\bibitem{Yagi:2005yb} 
  K.~Yagi, T.~Hatsuda and Y.~Miake,
  %``Quark-gluon plasma: From big bang to little bang,''
  Camb.\ Monogr.\ Part.\ Phys.\ Nucl.\ Phys.\ Cosmol.\  {\bf 23}, 1 (2005).
  %%CITATION = CMPCE,23,1;%%

%\cite{Adler:2002tq}
\bibitem{Adler:2002tq} 
  C.~Adler {\it et al.}  [STAR Collaboration],
  %``Disappearance of back-to-back high $p_{T}$ hadron correlations in central Au+Au collisions at $\sqrt{s_{NN}}$ = 200-GeV,''
  Phys.\ Rev.\ Lett.\  {\bf 90}, 082302 (2003).
  %[nucl-ex/0210033].
  %%CITATION = NUCL-EX/0210033;%%
  
%\cite{Schenke:2010rr}
\bibitem{Schenke:2010rr} 
  B.~Schenke, S.~Jeon and C.~Gale,
  %``Elliptic and triangular flow in event-by-event (3+1)D viscous hydrodynamics,''
  Phys.\ Rev.\ Lett.\  {\bf 106}, 042301 (2011).
  %[arXiv:1009.3244 [hep-ph]].
  %%CITATION = ARXIV:1009.3244;%%

%\cite{Danielewicz:1984ww}
\bibitem{Danielewicz:1984ww} 
  P.~Danielewicz and M.~Gyulassy,
 %``Dissipative Phenomena in Quark Gluon Plasmas,"
  Phys.\ Rev.\ D {\bf 31}, 53 (1985).
  %%CITATION = PHRVA,D31,53;%%

%\cite{Kovtun:2004de}
\bibitem{Kovtun:2004de} 
  P.~Kovtun, D.~T.~Son and A.~O.~Starinets,
  %``Viscosity in strongly interacting quantum field theories from black hole physics,''
  Phys.\ Rev.\ Lett.\  {\bf 94}, 111601 (2005).
  %[hep-th/0405231].
  %%CITATION = HEP-TH/0405231;%%

%\cite{Asakawa:2012yv}
\bibitem{Asakawa:2012yv} 
  M.~Asakawa, S.~A.~Bass and B.~M\"{u}ller,
  %``Center domains and their phenomenological consequences,''
  Phys.\ Rev.\ Lett.\  {\bf 110}, 202301 (2013).
  %[arXiv:1208.2426 [nucl-th]].
  %%CITATION = ARXIV:1208.2426;%%
  
%\cite{Adare:2008ab}
\bibitem{Adare:2008ab} 
  A.~Adare {\it et al.}  [PHENIX Collaboration],
  %``Enhanced production of direct photons in Au+Au collisions at $\sqrt{s_{NN}}=200$ GeV and implications for the initial temperature,''
  Phys.\ Rev.\ Lett.\  {\bf 104}, 132301 (2010).
  %[arXiv:0804.4168 [nucl-ex]].
  %%CITATION = ARXIV:0804.4168;%%
  
%\cite{Wilde:2012wc}
\bibitem{Wilde:2012wc} 
  M.~Wilde [ALICE Collaboration],
  %``Measurement of Direct Photons in pp and Pb-Pb Collisions with ALICE,''
  Nucl.\ Phys.\ A {\bf 904-905}, 573c (2013).
  %[arXiv:1210.5958 [hep-ex]].
  %%CITATION = ARXIV:1210.5958;%%
  
%\cite{McLerran:1993ni}
\bibitem{McLerran:1993ni} 
  L.~D.~McLerran and R.~Venugopalan,
  %``Computing quark and gluon distribution functions for very large nuclei,''
  Phys.\ Rev.\ D {\bf 49}, 2233 (1994);
  %[hep-ph/9309289].
  %%CITATION = HEP-PH/9309289;%%
  %
%\cite{McLerran:1993ka}
%\bibitem{McLerran:1993ka} 
  %L.~D.~McLerran and R.~Venugopalan,
  %``Gluon distribution functions for very large nuclei at small transverse momentum,''
  %Phys.\ Rev.\ 
  D {\bf 49}, 3352 (1994).
  %[hep-ph/9311205].
  %%CITATION = HEP-PH/9311205;%%
  
%\cite{Lappi:2006fp}
\bibitem{Lappi:2006fp} 
  T.~Lappi and L.~McLerran,
  %``Some features of the glasma,''
  Nucl.\ Phys.\ A {\bf 772}, 200 (2006).
  %[hep-ph/0602189].
  %%CITATION = HEP-PH/0602189;%%
  
%\cite{Mohapatra:2012ck}
\bibitem{Mohapatra:2012ck} 
  R.~K.~Mohapatra and A.~M.~Srivastava,
  %``Domain growth and fluctuations during quenched transition to QGP in relativistic heavy-ion collisions,''
  arXiv:1210.4718 [hep-ph].
  %%CITATION = ARXIV:1210.4718;%%
  
%\cite{Gross:1980br}
\bibitem{Gross:1981br} 
  D.~J.~Gross, R.~D.~Pisarski and L.~G.~Yaffe,
  %``QCD and Instantons at Finite Temperature,''
  Rev.\ Mod.\ Phys.\  {\bf 53}, 43 (1981).
  %%CITATION = RMPHA,53,43;%%
  
%\cite{Weiss:1980rj}
\bibitem{Weiss:1980rj} 
  N.~Weiss,
  %``The Effective Potential for the Order Parameter of Gauge Theories at Finite Temperature,''
  Phys.\ Rev.\ D {\bf 24}, 475 (1981).
  %%CITATION = PHRVA,D24,475;%%
  
%\cite{Belyaev:1991np}
\bibitem{Belyaev:1991np} 
  V.~M.~Belyaev, I.~I.~Kogan, G.~W.~Semenoff and N.~Weiss,
  %``Z(N) domains in gauge theories with fermions at high temperature,''
  Phys.\ Lett.\ B {\bf 277}, 331 (1992).
  %%CITATION = PHLTA,B277,331;%%
  
%\cite{Smilga:1993vb}
\bibitem{Smilga:1993vb} 
  A.~V.~Smilga,
  %``Are Z(N) bubbles really there?,''
  Annals Phys.\  {\bf 234}, 1 (1994).
  %%CITATION = APNYA,234,1;%%
  
%\cite{Deka:2010bc}
\bibitem{Deka:2010bc} 
  M.~Deka, S.~Digal and A.~P.~Mishra,
  %``Meta-stable States in Quark-Gluon Plasma,''
  Phys.\ Rev.\ D {\bf 85}, 114505 (2012).
  %[arXiv:1009.0739 [hep-lat]].
  %%CITATION = ARXIV:1009.0739;%%
  
%\cite{Borsanyi:2010cw}
\bibitem{Borsanyi:2010cw} 
  S.~Borsanyi, J.~Danzer, Z.~Fodor, C.~Gattringer and A.~Schmidt,
  %``Coherent center domains from local Polyakov loops,''
  J.\ Phys.\ Conf.\ Ser.\  {\bf 312}, 012005 (2011).
  %[arXiv:1007.5403 [hep-lat]].
  %%CITATION = ARXIV:1007.5403;%%
  
%\cite{Danzer:2010ge}
\bibitem{Danzer:2010ge} 
  J.~Danzer, C.~Gattringer, S.~Borsanyi and Z.~Fodor,
  %``Center clusters and their percolation properties in lattice QCD,''
  PoS LATTICE {\bf 2010}, 176 (2010).
  %[arXiv:1010.5073 [hep-lat]].
  %%CITATION = ARXIV:1010.5073;%%
  
%\cite{Lieb:1997mi}
\bibitem{Lieb:1997mi} 
  See for example, E.~H.~Lieb and J.~Yngvason,
  %``The Physics and mathematics of the second law of thermodynamics,''
  Phys.\ Rept.\  {\bf 310}, 1 (1999).
  %[cond-mat/9708200].
  %%CITATION = COND-MAT/9708200;%%
  
%\cite{Gale:2012rq}
\bibitem{Gale:2012rq} 
  C.~Gale, S.~Jeon, B.~Schenke, P.~Tribedy and R.~Venugopalan,
  %``Event-by-event anisotropic flow in heavy-ion collisions from combined Yang-Mills and viscous fluid dynamics,''
  Phys.\ Rev.\ Lett.\  {\bf 110}, 012302 (2013).
  %[arXiv:1209.6330 [nucl-th]].
  %%CITATION = ARXIV:1209.6330;%%
  
%\cite{CMS:2012aa}
\bibitem{CMS:2012aa} 
  S.~Chatrchyan {\it et al.}  [CMS Collaboration],
  %``Study of high-pT charged particle suppression in PbPb compared to $pp$ collisions at $\sqrt{s_{NN}}=2.76$ TeV,''
  Eur.\ Phys.\ J.\ C {\bf 72}, 1945 (2012).
  %[arXiv:1202.2554 [nucl-ex]].
  %%CITATION = ARXIV:1202.2554;%%
  
%\cite{Martinez:2010sc}
\bibitem{Martinez:2010sc} 
  M.~Martinez and M.~Strickland,
  %``Dissipative Dynamics of Highly Anisotropic Systems,''
  Nucl.\ Phys.\ A {\bf 848}, 183 (2010).
  %[arXiv:1007.0889 [nucl-th]].
  %%CITATION = ARXIV:1007.0889;%%
  
\end{thebibliography}

\end{document}